\documentstyle[12pt,psfig,epsf]{article}
\newcommand{\be}{\begin{equation}}
\newcommand{\ee}{\end{equation}}
\newcommand{\bef}{\begin{figure}}
\newcommand{\eef}{\end{figure}}
\newcommand{\bb}{\bibitem}

\newcommand{\nunu}{\nonumber}
\begin{document}
\vskip 0.2in
\begin{center}
{\Large {Relics of the Cosmological QCD Phase Transition}}
\vskip 0.25in
Abhijit Bhattacharyya$^a$, Jan-e Alam$^a$$\footnote{Present Address:
Physics Department, Kyoto University, Kyoto 606-8502, Japan}$, Sourav Sarkar$^a$, 
Pradip Roy$^a$,\\
\vskip 0.1in
Bikash Sinha$^{a,b}$, Sibaji Raha$^c$ and Pijushpani Bhattacharjee$^{d,e}$\\
\vskip 0.2in
\noindent
{\small {\it a) Variable Energy Cyclotron Centre,
     1/AF Bidhan Nagar, Calcutta 700 064, India}}
\vskip 0.1in
\noindent
{\small {\it b) Saha Institute of Nuclear Physics,
           1/AF Bidhan Nagar, Calcutta 700 064,
           India}}
\vskip 0.1in
\noindent
{\small {\it c) Department of Physics, 
	        Bose Institute, 
		93/1, A.P.C. Road,
                Calcutta 700 009,
           India}}
\vskip 0.1in
\noindent
{\small {\it d) Laboratory for High Energy Astrophysics, 
	        NASA/Goddard Space Flight Centre,\\ 
	        Code 661 Greenbelt, MD 20771, USA}}

\vskip 0.1in
\noindent
{\small {\it e) Indian Institute of Astrophysics,
		Bangalore - 560 034
           India}}
\end{center}
\addtolength{\baselineskip}{0.4\baselineskip} 

\parindent=20pt
\vskip 0.2in
\begin{abstract}
The abundance and size distribution of quark nuggets (QN), formed 
a few microseconds after the big bang due to first order QCD 
phase transition in the early universe, has been estimated. 
It appears that stable QNs could be a viable 
candidate for cosmological dark matter. The evolution of baryon 
inhomogeneity due to evaporated (unstable) QNs are also examined.
\end{abstract}
\vskip 0.1in
\section{Introduction}

As per the standard model, the universe, after a few micro seconds of
the big bang, underwent a phase transition from quarks and gluons to 
hadrons. There are well organised efforts to mimic such a phase 
transition in the laboratory through heavy ion collisions at 
ultra-relativistic energies~\cite{qm9697}. 
Although there are similarities between 
the two scenarios, the relatively long time scales in the early universe 
phase transition may indeed be more conducive to a reliable thermodynamic 
description. In the absence of any consensus on the order of QCD
phase transition for two light and one medium-heavy quarks form 
lattice calculations~\cite{lattice96} in the present work we 
assume an underlying picture of a first order phase transition
~\cite{witten} from quarks and gluons to hadrons. 
The central question we would like to address is what plausible remnants may 
have survived since that primordial epoch. Crawford and 
Schramm~\cite{crawford} and  Van Hove~\cite{van} argued that 
fluctuations in the horizon scale, triggered by the phase transition,
may lead to the formation of primordial black
holes, which could be as large as $M_\odot$, the solar mass. 
Schramm~\cite{schramm} has recently suggested 
that these black holes could even be the candidates
for the Massive Compact Halo Objects (MACHO's)~\cite{aubourg,alcock}, 
which had of late been discovered in the halo of the Milky way, in 
the direction of the Large Magellanic Cloud (LMC) using the 
gravitational lensing techniques. On the other hand, a first order 
QCD phase transition scenario
involving bubble nucleation at a critical temperature $T_c\sim 100-200$ MeV
could lead to the formation of quark nuggets (QN), made of
$u$, $d$ and $s$ quarks at a density somewhat larger than normal 
nuclear matter density. 
If these primordial QN's indeed survive till the present epoch, they could be a 
possible candidate for the baryonic component of the dark matter~\cite{witten}. 
Such a possibility would be
aesthetically rather pleasing, as it would not require invoking any 
exotic physics
nor would the success of the primordial (Big Bang) nucleosynthesis scenario
be materially affected~\cite{yang,schaeffer,rana,madsen}.

The central question in this context is, thus, whether the primordial QN's can
be stable on a cosmological time scale.  The first study  on this issue 
was addressed by Alcock and Farhi~\cite{alcock1} who argued that neutrons 
could be liberated and emitted from a QN at temperature $T\geq I_N$, where
$I_N$ ($\sim$ 20-80 MeV) is the neutron "binding energy" {\it i.e.}
the difference 
between energy per baryon in strange matter at zero temperature 
and the mass of the nucleon. They calculated the
rate of baryon evaporation from QNs by using detailed balance
between the two processes of neutron absorption and emission in a system
consisting of QNs (of all possible baryon numbers) and neutrons. They
also assumed a geometric cross section for the neutron absorption
by the QNs and complete transparency of the nugget surface to neutrons.
This gave the result that QNs even with the largest allowed initial
baryon number (i.e. the total baryon number $N_{B,hor}\approx 10^{49}
(100 MeV/T)^3$ contained in a horizon-size volume of the universe,
at the time of formation of the nugget, $T$ being the temperature of the
universe at that time) were unstable with respect to baryon evaporation.
These results apparently eliminated the possibility of any QN surviving
till the present epoch.

Madsen et al~\cite{madsen1} then 
pointed out that, since evaporation was a surface process, emission 
of neutrons (proton emission was suppressed
due to the Coulomb barrier) made the surface layer deficient in $u$ and $d$ 
quarks, although relatively enriched in $s$ quarks.
The rate of conversion of $s$ quarks back to $u$ and $d$ quarks 
as well as convection of $u$ and $d$ quarks from
the core of the nugget to the surface were both too slow to
establish flavor chemical equilibrium between $u$, $d$ and
$s$ quarks on the surface layer. 
As a result of this deficiency
of $u$ and $d$ quarks in the surface layer, further nucleon evaporation was
suppressed. Madsen et al found that
QNs with initial baryon number $N_B\ge 10^{46}$ could well be stable against
baryon evaporation.

QNs at temperatures above a few MeV could also be subject to the process of
"boiling"~\cite{alcock2}, {\it i.e.}, spontaneous nucleation of 
hadronic bubbles in the bulk
of the nugget and consequent conversion of the nugget into nucleons. Madsen
and Olesen~\cite{madsen2} however, 
showed that although boiling is thermodynamically
allowed, the time scale for bubble nucleation inside QNs is too long
compared to the time scale of surface evaporation for reasonable values
of the parameters used. 

All of the above studies used thermodynamic and binding energy arguments
to calculate the baryon evaporation rate; the microscopic dynamics 
determining the probability
of baryon formation and emission had been neglected. 
Clearly, for a realistic description of
the process a dynamical model of baryon emission from QNs was needed.

Earlier, Bhattacharjee {\it et al}~\cite{pijush} 
used the chromoelectric flux tube model (inspired by QCD) to
demonstrate that the QN's
would survive against baryon evaporation, if the baryon number of the quark
matter inside the nugget was larger than $10^{42}$. For reasons explained in
ref.~\cite{pijush}, this estimate is rather 
conservative. Sumiyoshi and Kajino~\cite{sumi}
estimated within a similar approach that a 
QN with an initial baryon number
$\sim 10^{39}$ would survive against baryon evaporation. 

It may also be mentioned in this context that Madsen and 
Riisager~\cite{madsen} had calculated the minimum
size of the QNs from the constraint imposed by the primordial
$He$ abundance. The authors showed that the minimum radius of the
QNs should be more than $\sim 10^{-6}$ cm, otherwise the absorption 
of neutrons by the QNs will upset the neutron to proton 
ratio in the universe, resulting in a helium 
abundance in complete disagreement with the experimental results. 

In spite of these efforts, not much attention has been paid  
towards the issues of formation and size-distribution of the surviving quark 
nuggets in the universe. The size-distribution and
abundance of the QNs is very important in the context of their 
candidature as dark matter. The calculation of a lower cut-off in size,
if any, would 
tell us what can be the minimum size and the baryon number content
of a QN that we should  look for. On the other hand, the distribution
function also indicates the most probable size of the QNs.  
Such studies in the framework of the 
GUT (Grand Unified Theory) phase transition and the associated "Trapped
False Vacuum Domains" (TFVD) have been done earlier
~\cite{kodama}.  The basic contention of this paper is to carry out
these studies for the QCD phase transition in the early universe.
We also study the evolution of baryon inhomogeneity created by evaporating
QNs, with baryon number lower than the survivability window, as 
mentioned above, due to 
the conduction of heat by neutrinos, and consequently the dissipation 
of the baryon inhomogeneities.  

We organise the paper as follows. In the next section we evaluate 
the size-distribution of the QNs. The results for different nucleation
rates have been discussed. Section 3  contains the evolution 
of baryon inhomogeneity which originated due to the unstable QNs. 
In section 4 we conclude.

\section{The size-distribution of quark nuggets}

The evolution of the universe during the QCD phase transition is
governed by Einstein's equations, 
\be
\left(\frac{\dot R}{R}\right)^2=\frac{8\pi\rho}{3m_{pl}^2};~~~~
\frac{d(\rho R^3)}{dt}+P\frac{dR^3}{dt}=0
\label{rob}
\ee
where $\rho$ is the energy density, $P$ the pressure and $m_{pl}$ 
the Planck mass. In the above equation, $R$ is the cosmological scale
factor in the Robertson-Walker space time and is defined by the relation 
\begin{eqnarray}
ds^2 &=& -dt^2 + R^2 dx^2 = R^2(-d\xi^2 + dx^2) \nonumber\\
dx^2 &=& dX^2 + X^2(sin^2\theta d\phi^2 + d\theta^2)
\end{eqnarray}
where $X$ is the co-ordinate 
radius {\it i.e.} the radius in the unit of the cosmological scale 
factor $R(t)$. 

It is well known that in a first order phase transition, 
the quark and the hadron phases co-exist in a mixed phase at the
critical temperature of transition. Around the 
critical temperature, the universe consists of leptons, photons and the 
massless quarks, anti-quarks and gluons, described in our case by the MIT 
bag model with
an effective degeneracy $g_q (\sim 51.25)$. 
The hadronic phase contains relativistic $\pi$-mesons, photons and
leptons with a small baryon content ($\rho_B/\rho_\gamma \sim 
10^{-10}$) and is described by an equation of state corresponding to
massless particles with an effective degeneracy $g_h=17.25$. 
The energy densities and the pressures of hadronic and quark matter
are given by
\begin{eqnarray}
\rho_h &=& {{\pi^2 g_h} \over 30} T^4; \,\,\,\,\,\,
\rho_q = {{\pi^2 g_q} \over 30} T^4 +B\nonumber\\
P_h &=& {{\pi^2 g_h} \over 90} T^4; \,\,\,\,\,\,
P_q = {{\pi^2 g_q} \over 90} T^4 -B
\end{eqnarray}
where $B$ is the bag constant.

The evolution of the scale factor in the mixed phase is given by 
(see also \cite{fuller}),
\be
R(t)/R(t_i)=\left[cos\left(arctan\sqrt{3r}-
\sqrt{\frac{3}{r-1}}(t-t_i)/t_c\right)\right]^{2/3}/
\left[cos\left(arctan\sqrt{3r}\right)\right]^{2/3}
\ee
In the process we also get the volume fraction of the quark matter 
$f(t)$ in the mixed phase as
\be
f(t)=\frac{1}{3(r-1)}\left[tan\lbrace arctan\sqrt{3r}-
\sqrt{\frac{3}{r-1}}\frac{t-t_i}{t_c}\rbrace\right]^2
-\frac{1}{r-1}
\ee
where $r \equiv \rho_q/\rho_h$, $t_c = \sqrt{3m_{pl}^2/8\pi B}$~is the
characteristic time scale for the QCD phase transition in the early 
universe and $t_i$ is the time when phase transition starts. (The
definition of $r$ and hence the expressions of $R(t)$ and $f(t)$ here
are somewhat different from that in ref. \cite{fuller}).
The characteristic time scale depends on the 
bag constant and hence on the critical temperature of the quark-hadron 
phase transition ($T_c$). In fact $t_c = 144 \mu s$ for $T_c = 100 MeV$
and $t_c = 64 \mu s $ for $T_c = 150 MeV$. 

In the coexisting phase, the temperature of the universe remains constant at
$T_c$, the cooling due to expansion being compensated by the liberation of
the latent heat. In the usual picture of bubble nucleation in a first order
phase transition scenario hadronic matter starts appearing 
as individual bubbles. With the progress of time, more and more hadronic
bubbles form, coalesce and eventually percolate to form an infinite network 
of hadronic matter which traps the quark matter phase into finite domains. 
The time when the percolation takes place is usually referred to as the 
percolation time $t_p$, determined by a critical volume fraction 
$f_c$, ($f_c \equiv f(t_p)$) of the quark phase.

In an ideal first order phase transition, the fraction of the high
temperature phase decreases from the critical value $f_c$, as these domains
shrink. For the QCD phase transition, however, these domains could become
QN's and as such, we may assume that the lifetime of the
mixed phase $t_f\sim t_p$.

As mentioned above, just after percolation one can have pockets of 
quark matter trapped as bubbles in the ambient hadronic matter. The 
probability that
a spherical region of co-ordinate radius $X$ at time $t_p$ with nucleation 
rate $I(t)$ lies completely within the quark matter domain is given by 
\cite{kodama},
\be
P(X,t_p)=\exp\left[-\frac{4\pi}{3}\int_{t_i}^{t_p}dtI(t)R^3(t)\left(X
+X(t_p,t)\right)^3\right]
\ee
where $X(t_p;t)$ is the coordinate radius of a bubble, at time $t_p$, which 
was nucleated at time $t$. 

For convenience, let us now define a new set of variables 
$z=X R(t_i)/vt_c$, $x=t/t_c$ and $r(x)=R(x)/R(x_i)$; where $v$ is the radial 
growth velocity of the nucleating bubbles. Then 

\be
P(z,x_p)=\exp\left[-\frac{4\pi}{3}v^3t_c^4\int_{x_i}^{x_p}dxI(x)\left(zr(x)
+y(x_p,x)\right)^3\right]
\ee
where
\be
y(x,x\prime)=\int_{x\prime}^x{r(x\prime)}/{r(x\prime\prime)}
dx\prime\prime
\ee
So the fraction of quark matter present at time $t_p$ is 
\be
f_c = P(0,x_p)=\exp\left[-\frac{4\pi}{3}v^3t_c^4\int_{x_i}^{x_p}dxI(x)
y^3(x_p,x)\right]
\ee

Let us now look at the size-distribution of the 
Trapped Quark Matter Domain (TQMD). 
In order to do so we will follow the procedure of ref. \cite{kodama}.
The difference of our work from that of Kodama {\it et al.} 
is that we have considered exactly
spherical nuggets whereas they have included a deformation factor. It
should however be noted that the deformation factor, as found by Kodama
{\it et al}, is small. Moreover, due to the presence of non-zero surface
tension in case of QCD phase transition the bubbles are likely to be 
spherical. Even more
importantly, we focus our attention to the percolation time $t_p$ when
the hadronic matter forms the ambient background. All these
considerations allow us to consider the false vacuum domains (the quark
phase) as being spherical in shape. Following Ref.~\cite{kodama} let us 
assume that $F(X;t) dX$ is the number of TQMDs per unit volume 
within the size $\left\{X,X+dX\right\}$ at time $t$. Then $P(X,t)$ can
be thought to be the probability that a QN of coordinate radius
$X$ at a fixed position is contained in a  
TQMD. Now a TQMD of size $\eta$ can contain such a 
sphere of size $X$ only when the center of TQMD lies within the
coordinate radius $\eta - X$ from the center of the sphere. 
If $\alpha$ is the minimum size of a TQMD {\it i.e.} $F(X,t)$ vanishes
for $X < \alpha$ then one can write 
\be
P(X;t) = \int_{\alpha+X}^{\infty} {{4\pi} \over 3} (\eta - X)^3 F(\eta;t) d\eta
\ee

The distribution function vanishes for $X < \alpha$. One can now 
solve the above equation using Laplace transformation to obtain 
$F(X)$ (see appendix).

The result, in terms of $z$, is
\begin{eqnarray}
F(z) &=& {{3 {\hskip 0.04in} \theta(z-\alpha) R(t_i)^4} \over 
{4 \pi \alpha^3 v^4 t_c^4}} 
\left[-P'(X-\alpha) -{{3P(X-\alpha)} \over \alpha}\right. \nonumber\\
&+& \left.{1 \over \alpha^2} \int_0^\infty d\eta P(\eta+X-\alpha) 
\left\{\lambda e^{(-\lambda \eta/\alpha)} 
+\omega e^{(-\omega \eta/\alpha)} 
+{\bar{\omega}} e^{(-\omega \eta/\alpha)}\right\}\right]\nonumber\\
&=& {{R(t_i)^4} \over {v^4 t_c^4}} f(z) 
\end{eqnarray}

The solution of the equation $F(\alpha) = 0$ gives the minimum size of
the quark-nugget $\alpha$. Now, the number of nuggets per unit volume is
given by 
\begin{eqnarray}
n_Q &=& R^{-3}(t_p) \int_\alpha^\infty F(X) dX \nonumber\\
 &=& R^{-3}(t_p) \int_\alpha^\infty {{R^3(t_i)} \over {v^3 t_c^3}} 
 f(z) dz
\end{eqnarray} 

The volume of each quark nugget is given by 
${4 \over 3} \pi (zvt_c)^3$. Since the visible baryon constitutes only
ten per cent of the closure density
($\Omega_B=0.1$ from standard big bang 
nucleosynthesis), a total of $10^{50}$ baryons will
close the universe baryonically at $T=100$ MeV. We emphasize at this point that
these QNs would not disturb the standard primordial nucleosynthesis
results to any considerable extent, as they would not participate in
usual nuclear reactions.
Therefore, if we assume that the total baryon content of the dark
matter is carried by the quark nuggets, then,
\be
N_B = 10^{50} (100/T (MeV))^3 = V_H {{4\pi R^3(t_i)} \over {3
R^3(t_p)}}\rho_B \int_\alpha^\infty f(z) z^3 dz
\ee
where $V_H$ is the horizon volume and $\rho_B$ is the baryon density
inside each nugget. We now solve the above equations self-consistently 
to obtain $\alpha$, $t_p$ and $f_c$. These values are then used to study
the size-distribution of the quark nuggets. 
 
To calculate the size and distribution of QNs we need to know
the rate of nucleation during the phase transition process.
Many authors~\cite{cottingham,kapusta} have proposed various nucleation rates
for the QCD phase transition.
In the absence of a consensus as to which rate is preferable to others,
we look at several of them. We begin with the rate proposed by Cottingham {\it et.al.} \cite{cottingham} 
which is based on the 
Lee-Wick model of effective QCD. They calculated the Lee-Wick potential at finite
temperature to  obtain the following 
nucleation rate  
\be
I(t) = T^4 \left({S \over {2\pi T}}\right)^{3/2}\exp\left(-S/T\right) 
\ee
with
\begin{eqnarray}
S &=& {{2\pi} \over 3} \left({{m\sigma_0^2} \over 3}\right)^3 {1 \over P^2} 
\nonumber\\
P &=& {{7\pi^2} \over 30} \left(T_c^4 - T^4\right)
\end{eqnarray}
where $\sigma_0 = 100 MeV$ and $m = 939 MeV$.  

Csernai and Kapusta~\cite{kapusta} proposed a nucleation rate which is of the 
form 
\be
I(t) = r_T\exp\left[-\Delta/(1-t/t_c)^2\right] 
\ee
with
\be
\Delta = {{16\pi \sigma^3} \over {48B^2T_c}}  
\ee
where $\sigma$ is the surface tension and $r_T$ is a temperature
dependent constant. They have calculated the pre-exponential factor 
$r_T$ from an effective field theory of QCD. This nucleation rate 
is the same as the general form proposed by Landau and 
Lifshitz~\cite{landau}, apart from the pre-exponential factor.

\begin{table}
\begin{center}
\begin{tabular}{|cccc|}
\hline
          &         &       &    \\
  $\sigma$&    $T_c$&minimum radius &$N_{QN}$  \\
(MeV/fm$^{-2}$)&   (MeV) &(meter)&     \\
          &         &       &    \\
\hline
          &          &       & \\
Cottingham {\it et. al.}    &100 & 1.66  & 7420 \\
          &        &       & \\
    &150  & 0.083  & $1.7 \times 10^7$ \\
          &        &       & \\
\hline
10 &     &       & \\
Csernai {\it et. al.}&100&0.117& $2.1 \times 10^{7}$ \\
   &                      &    &    \\
   & 150 &0.0096& $3.8 \times 10^{10}$ \\
   &     &    &    \\
\hline
50 &     &    & \\
Csernai {\it et. al.}& 100 & 1.25& $1.7 \times 10^4$ \\
   &     &    &    \\
   & 150 &0.0882& $1.4 \times 10^7$ \\
   &     &    &    \\
\hline
\end{tabular}
\end{center}
\caption{ Different values of minimum radius and $N_B$ for the different 
nucleation rates.}
\end{table}

Let us discuss the results obtained so far. 
In table I we have shown the dependence of the minimum radius of a
quark nugget and the number of QNs within the horizon just after the QCD
phase transition on the value of $T_c$ for different nucleation rates.  
For the nucleation rate proposed by Csernai and Kapusta, we have varied 
the surface tension from $\sigma = 10-50 MeV fm^{-2}$. We have found
that the minimum radius varies from $9.6 \times 10^{-3}$ meter to $1.66$
meter. 

In fig.~\ref{cot} we have plotted the distribution of QN, $f({\bar n}_B)$, as a 
function of ${\bar n}_B$ using the nucleation rate proposed by Cottingham 
{\it et.al.}, for different values of $T_c$, where ${\bar n}_B$ is the baryon
number content of a single QN. We see that for $T_c = 100 MeV$
distribution of QN peaks at baryon number $\sim 7 \times 10^{45}$ and there is 
almost no QN with baryon number larger than  $10^{47}$. For $T_c = 150 MeV$ 
these values are $10^{42}$ and $10^{43}$ respectively. 
In figs.~\ref{10kap}-\ref{50kap}, 
similar results have been shown for nucleation rate proposed by 
Csernai and Kapusta, for $\sigma = 10$ and $50 MeV fm^{-2}$, respectively. 
We see that for a fixed $T_c$, the minimum radius increases with 
increase in $\sigma$. Figure 3 shows that 
the maximum number of nuggets are around a baryon number of 
$\sim 2.5 \times 10^{42}$ 
at $T_c = 100 MeV$ whereas the number of nuggets goes to zero after 
$ 4 \times 10^{43}$. Similar results for $T_c = 150 MeV$ are also shown 
in the figures.
We should mention at this stage that the lower cutoffs that we have 
obtained here are certainly allowed by the study of Madsen {\it et.al.} 
\cite{madsen} within the reasonable set of parameters. 

The question which arises next is whether these nuggets
will survive till the present epoch.
Earlier studies \cite{pijush} have shown that the nuggets having baryon
number less than $10^{42}$ will not survive till the present epoch. 
This suggests that all the cases
considered here for $T_c = 100 MeV$ will give stable nuggets. However,
from figure 1 and 3, it can be seen that, for $T_c = 150 MeV$, some
of the nuggets will not survive. Fig. 2. suggests that for $\sigma = 10
MeV fm^{-2}$, none of the nuggets will survive when $T_c = 150 MeV$.
Given the present state of the art, there is no way to choose any one of the
possibilities as the preferred one. We should therefore consider the
situation that while some nuggets may indeed be stable and constitute
cold dark matter, some smaller nuggets may evaporate, creating sizeable
baryon inhomogeneities. In the next section we will study the evolution of 
these inhomogeneities with time/temperature.

\section{Evolution of baryon inhomogeneities due to evaporated quark nuggets}

Our aim in this section is to study the implications of those QNs that
do evaporate away, {\it assuming}, of course, that they were formed in the
early universe. When a QN dissociates into nucleons,
the latter initially form a clump with high baryon overdensity relative to
the density
of baryons in the ambient universe. The baryon density in the clump then
gradually decreases as various physical processes tend to `flatten' the
clump. We study the evolution of the highly non-linear baryonic
inhomogeneities represented by these high density clumps due to
dissociated QNs created after
the epoch of quark-hadron phase transition at $T\sim 100 MeV$, $T$ being the
temperature of the universe. 

The evolution of large, non-linear baryon inhomogeneities in the early
universe has been studied in detail recently, especially 
in the context of possible creation through electroweak baryogenesis
process \cite{heckler,jedamzik,jedam1}
of large baryon inhomogeneities during the epoch of a possible first-order
electroweak symmetry-breaking phase transition at $T\sim 100 GeV$.
The single most
dominant physical process that determines the evolution of large baryon
inhomogeneities in the early universe before the epoch of neutrino decoupling
(at $T\sim 1 MeV$) is the so-called ``neutrino inflation''. Any large
baryonic clump in pressure equilibrium
with the ambient universe would have a slightly lower temperature inside the
clump relative to the temperature of the ambient surroundings, due to the 
excess pressure contributed by the excess baryons inside the clump.
As a result, heat would be conducted into the
clump from the ambient medium. The particles most efficient in conducting heat
into the clump are the neutrinos which have, by far, the largest mean free
path (MFP) amongst all the relevant elementary particles. As neutrinos cross the
clump they
deposit energy into the clump thereby heating up the clump. The clump then
expands in order to
achieve pressure equilibrium under this changed circumstance, and so the
baryon density within the clump decreases as the clump expands. This 
process of expansion
(``inflation'') of the clump due to neutrino heat conduction continues until
the neutrinos decouple at around $T\sim 1 MeV$. For any given size of
a clump, the time scale on which a
clump achieves pressure equilibrium with the surroundings is essentially the
hydrodynamic expansion time scale or the time 
taken by sound to traverse the clump,
which can be shown to be smaller than the heat transport time scale for
neutrinos. It is, therefore, a good approximation to treat the
evolution of the clump as going through a succession of pressure equilibrium
stages with decreasing density inside the clump. After neutrino decoupling
the evolution of the clump is determined mainly by the process of baryons
slowly diffusing out of the high-density clump to the ambient medium.

In this section, we study
the evolution of the baryonic clumps created by evaporated QNs
under neutrino inflation.
It is to be mentioned here that the linear
relationship, assumed in ref.\cite{heckler}, between the
baryon overdensity within a clump and the
fractional temperature difference of the clump relative to the ambient
temperature, turns out to be invalid in our case of extremely large initial
baryon overdensity created by the evaporated QNs, as we discuss below.
As a consequence, we need to numerically solve the full non-linear pressure
equilibrium equation for a clump in order to obtain the relationship between
those quantities. Furthermore, the initial baryon overdensity within the clump
in our case can be so large (e.g. $\sim10^{12}$)\cite{iso}
 that baryon-to-entropy ratio
within the clump could be initially greater than unity
in which case the dominant contribution to the MFP of neutrinos would come
initially from neutrino-nucleon
scattering rather than from neutrino-lepton scattering assumed in
ref.\cite{heckler}. The above two considerations make a
straightforward application of the results of ref.\cite{heckler}
 invalid in our
case of large baryonic
inhomogeneities due to evaporating QNs; hence the need to do an {\it ab initio}
calculation for inhomogeneities with initial overdensity significantly larger
than those studied in ref.\cite{heckler}. In this respect, we
believe the calculations in this paper, although done in the specific
context of inhomogeneities due to quark nuggets, have much wider validity
and application. We would like to mention at this stage that we are
interested to study only the neutrino inflation process which will be
operative till $T = 1 MeV$. So we restrict our calculation upto that
temperature. Also, the only difference from the work of ref. \cite{heckler}
is that we have solved the full non-linear pressure equation.

The pressure equilibrium equation for a baryonic clump with baryon number
density $\rho_B^*\equiv\delta_N\rho_B$ and temperature
$T^*\equiv T(1-\delta T)$ in the background
universe at temperature $T$ and baryon number density $\rho_B$ can be
written as
\be
\rho_B^*\,\,T^*+{1\over3}g_{eff}(T^*)aT^{*4}=\rho_B T+ {1\over3}g_{eff}
(T)aT^4
\label{a1}
\ee
where $g_{eff}(T)$ is the effective number of relativistic degrees of
freedom
in the universe at temperature $T$ contributing to the energy
density and pressure, and $a=\pi^2/30$. The baryons within
the clump as well as outside are assumed to be ideal gases of non-relativistic
particles with pressures $\rho_B^*T^*$ and $\rho_B T$, respectively.
Assuming $\delta_N\ge 1$, $\delta T\le 1$, and
$g_{eff}(T^*)\approx g_{eff}(T)$, we get from eq.(~\ref{a1}),
\be
\delta T\simeq {\eta\delta_N\over 1+\eta\delta_N}
\label{a2}
\ee
where $\eta=\rho_B/s$, $s$ being the entropy density.
The baryon-to-entropy ratio $\eta$ in the universe is essentially constant
for the temperature range of our interest, $s\approx2.6\times10^8
\Omega_B^{-1}h^{-2}$, where $\Omega_B$ is the baryonic mass density in the
universe in units of the closure density, and $h=H_0/(100
km,sec^{-1},mpc^{-1})$, $H_0$ being the present value of the Hubble
constant.

Eq.(\ref{a2})shows that if the baryon overdensity $\delta_N$ in the clump 
satisfies
the condition $\eta\delta_N\ll 1$, then $\delta T\simeq\eta\delta_N$, i.e.,
$\delta T$ is linearly proportional to $\delta_N$~\cite{heckler}. 
On the other hand, for
overdensities satisfying $\eta\delta_N\gg 1$, Eq.(\ref{a2}) gives 
$\delta T\sim 1$,
which is inconsistent with the assumption $\delta T\ll 1$ under which 
Eq.(\ref{a2})
is derived. Clearly, then, for sufficiently large overdensities for which
$\delta_N\ge\eta^{-1}$, the assumption $\delta T\ll 1$ is not valid, and so
we need to solve the full non-linear pressure equilibrium equation, 
Eq.(\ref{a1}),
to obtain the relationship between $\delta T$ and $\delta_N$. This is
the essential difference between our work and that of
ref.\cite{heckler}. The result is demonstrated in figure \ref{inhom}. 

Now, for a given overdensity $\delta_N$ of the clump at some time $t$ when the
temperature of the universe is $T$, the rate of energy deposited into the
clump by neutrinos depends upon whether the size $L$ ($=2R$, $R$ being the
radius) of the
clump (assumed spherical) is larger or smaller than the MFP ($\lambda_{\nu}$),
of neutrinos through the clump at that time. For $\lambda_{\nu}\le L$
the clump will not inflate by any significant amount because
the energy deposition by ambient neutrinos will occur mainly in a thin
surface layer of the clump leaving the bulk of the clump unaffected. Indeed
in this case the heating of the clump will be governed by slow diffusion 
~\cite{jedamzik} of
neutrinos inside the clump. However, $\lambda_\nu$ in the early universe
increases rapidly as T decreases, typically, $\lambda_\nu\propto T^{-5}$. This
means that a clump of any given size $L$ will quickly come within the
``neutrino horizon'', such that  $\lambda_\nu$ becomes larger than $ L$ 
before any
significant neutrino inflation of the clump takes place. Indeed, most of the
neutrino inflation of the clump will take place when
$L\sim\lambda_{\nu}$ and $L\ge\lambda_{\nu}$. The evolution of
$\delta_N$ with time (temperature) 
is governed by the following differential equations~\cite{heckler,puri}
\be
{d\delta_N\over dt}=-{4\over R}{\rho_\nu\over\rho}\delta T\delta_N
\label{a3}
\ee

for $L\sim \lambda_\nu$,
\be
{d\delta_N\over dt}=-{3\over 4}
{1\over\lambda_\nu}{\rho_\nu\over\rho}\delta T\delta_N
\label{a4}
\ee

for $L>\lambda_\nu$.

\vskip 0.1in

The typical values of the overdensity $\delta_N$ and size $L$ of those
overdensities expected from QN evaporation, are calculated by Iso et al
\cite{iso}. 
The values of $\delta_N$ could be as large as $10^{12}$ and $R\sim 10$ cm.
Since nothing is known about the initial overdensity $\delta_N$ and the
length scale $L$, we study the evolution of the baryon overdensities 
for various initial values
of $\delta_N$ and $L$ by solving equation (\ref{a3}) and (\ref{a4}).
In figure \ref{inhom} we have shown the importance of considering the
non-linear term. It can be seen from the figure that at high $\delta T
/T$ the linear relation breaks down quite substantially.
The nuggets which will not be stable against evaporation will form 
highly dense baryonic lumps. We have studied the evolution of these
lumps with time. Neutrinos play an important role in the
evolution of these lumps up to $1$ MeV. The results are shown in 
fig.~\ref{infl}. Lumps with initial overdensity
$\le 10^8$ is not affected by the neutrino conduction.
The final values of the overdensity is  smaller
for higher initial temperature. 

From the above discussion and fig. 5 it is clear that some 
overdensity is left out after the neutrino inflation which is of the
order of $10^7$. This overdensity, as it looks, is a sizeable amount.
The baryon diffusion starts dominating after the neutrinos fall out of
equilibrium ($T \sim 1 MeV$). From $T = 1 MeV$ to the beginning of the
nucleosynthesis {\it i.e.} $T = 0.1 MeV$ baryon diffusion is the most
dominant process as far as the dissipation of the overdensities are
concerned. If the baryon diffusion lengths are larger than the
typical size of the inhomogeinities then the overdensities will
be washed out due to this process. As a result these objects will not
alter the standard big bang nucleosynthesis scenario.
This has been shown  for the baryon inhomogeinity
created at the Electro-Weak scale by 
Brandenberger {\it et. al.} \cite{bran}. 
These findings once
again supports the existence of nuggets. If the evaporating nuggets
would have left very high asymmetries in the universe then the observed
$He^4$ abundance, which is thought be very well determined, would have
been violated, a scenario not very comfortable with the survival of
quark nuggets. 

\section{Conclusion}
In this work we have estimated the abundance of quark nuggets in 
various nucleation scenarios with different values of critical
temperature and  surface tension of the bubble. We have found 
that within a reasonable set of parameters QNs may be a possible
candidate for cosmological dark matter. The evolution of 
baryon inhomogeneities, formed due to the unstable QNs have
also been studied.

\section{Appendix}
In this appendix we follow Ref.~\cite{kodama} to solve the following 
integral equation,
\be
P(X;t) = \int_{\alpha+X}^{\infty} {{4\pi} \over 3} (\eta - X)^3 F(\eta;t) 
d\eta
 \ee
Differentiating the above equation for four times we get
\be
{3 \over {4\pi}} P^{(4)} (X) = -D(\alpha \partial_X) \,\,F(X)
\ee
where 
\be
D(\alpha \partial_X) = \alpha^3 {{\partial^3} \over {\partial X^3}} 
-3\alpha^2 {{\partial^2} \over {\partial X^2}} 
-6\alpha {{\partial} \over {\partial X}} - 6
\ee
Let 
\be
{\cal L} \left(F(X)\right) = {\bar F}(p) = \int_0^\infty F(X) 
e^{-pX} dp
\ee
where ${\cal L}(A)$ is the Laplace transform of $A$. Now using the 
Laplace transformations of the derivatives we have
\be
{3 \over {4\pi}} P^{(4)} (X) = D(\alpha p) {\bar F}(p)e^{-\alpha p}
\ee
where we have neglected an arbitrary constant.
Now,

\begin{eqnarray}
F(X) &=& {3 \over {4\pi}} {1 \over {2\pi i}} \int_{c-i\infty}^{c+i\infty}
{{P^{(4)}(p)} \over {D(\alpha p)}} e^{p(X-\alpha)} dp \nonumber\\
&=& {3 \over {4\pi}} \int_0^\infty d\eta P^{(4)}(\eta) 
{1 \over {2\pi i}} \int_{c-i\infty}^{c+i\infty} {{e^{p(X-\alpha -\eta)}}
\over {D(\alpha p)}}  dp \nonumber\\
&\times&\left[-P^{(1)}(X-\alpha) - {3 \over {\alpha}}P(X-\alpha) \right.\nunu\\
& &+{1 \over {\alpha^2}} 
\int_0^\infty d\eta P(\eta+X-\alpha) \left\{\lambda e^{-\lambda \eta /\alpha
}\right.\nunu\\
& &\left. \left.+ \omega e^{-\omega \eta /\alpha}
+ {\bar \omega} e^{-{\bar \omega} \eta /\alpha}\right\} \right]
\end{eqnarray}

\begin{eqnarray}
F(X) &=& {{3 {\hskip 0.04in} \theta(X-\alpha)} \over {4 \pi \alpha^3}} 
\left[-P'(X-\alpha) -{{3P(X-\alpha)} \over \alpha}\right. \nonumber\\
&+& \left.{1 \over \alpha^2} \int_0^\infty d\eta P(\eta+X-\alpha) 
\left\{\lambda e^{(-\lambda \eta/\alpha)} 
+\omega e^{(-\omega \eta/\alpha)} 
+{\bar{\omega}} e^{(-\omega \eta/\alpha)}\right\}\right]
\label{apeq}
\end{eqnarray}
where $\lambda$, $\omega$ and ${\bar {\omega}}$ are the solutions 
of the equation 
\be
x^3 - 3x^2 +6x - 6 = 0
\ee

In terms of the variable $z$  Eq.~(\ref{apeq}) looks like

\begin{eqnarray}
F(z) &=& {{3 {\hskip 0.04in} \theta(z-\alpha) R(t_i)^4} \over 
{4 \pi \alpha^3 v^4 t_c^4}} 
\left[-P'(X-\alpha) -{{3P(X-\alpha)} \over \alpha}\right. \nonumber\\
&+& \left.{1 \over \alpha^2} \int_0^\infty d\eta P(\eta+X-\alpha) 
\left\{\lambda e^{(-\lambda \eta/\alpha)} 
+\omega e^{(-\omega \eta/\alpha)} 
+{\bar{\omega}} e^{(-\omega \eta/\alpha)}\right\}\right]\nonumber\\
&=& {{R(t_i)^4} \over {v^4 t_c^4}} f(z) 
\end{eqnarray}
with 
\begin{eqnarray}
\alpha \rightarrow {{\alpha v t_c} \over {R(t_i)}}; {\hskip 0.14in} 
\eta \rightarrow {{\eta v t_c} \over {R(t_i)}} 
\end{eqnarray}

\newpage

\bef[htb]
\centerline{\psfig{figure=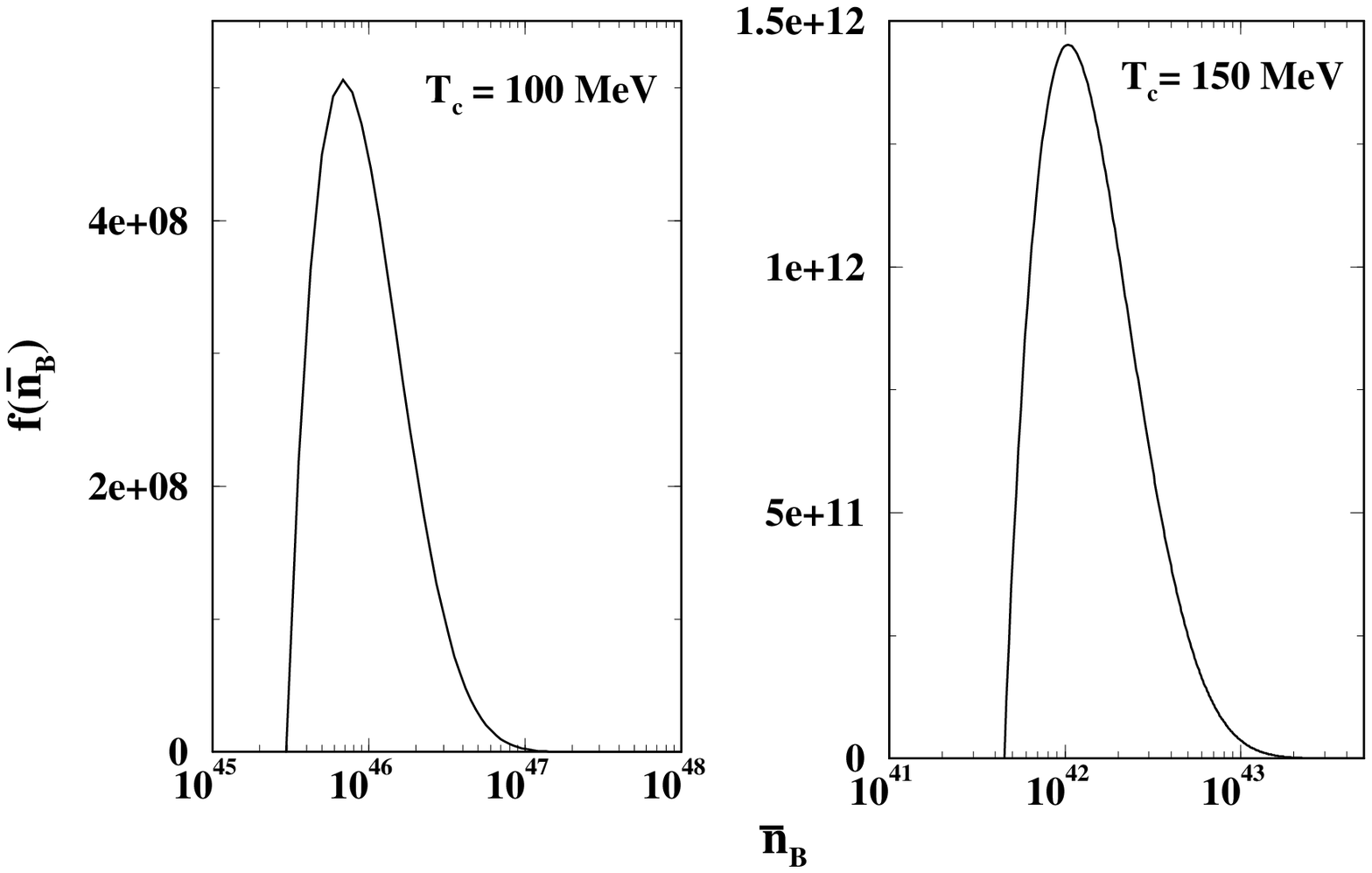,height=8cm,width=15cm}}
\caption{Distribution of QN, $f({\bar n}_B)$, as a function of ${\bar n}_B$ 
using nucleation rate proposed by Cottingham {\it et. al.} }
\label{cot}
\eef
\newpage
\bef[hb]
\centerline{\psfig{figure=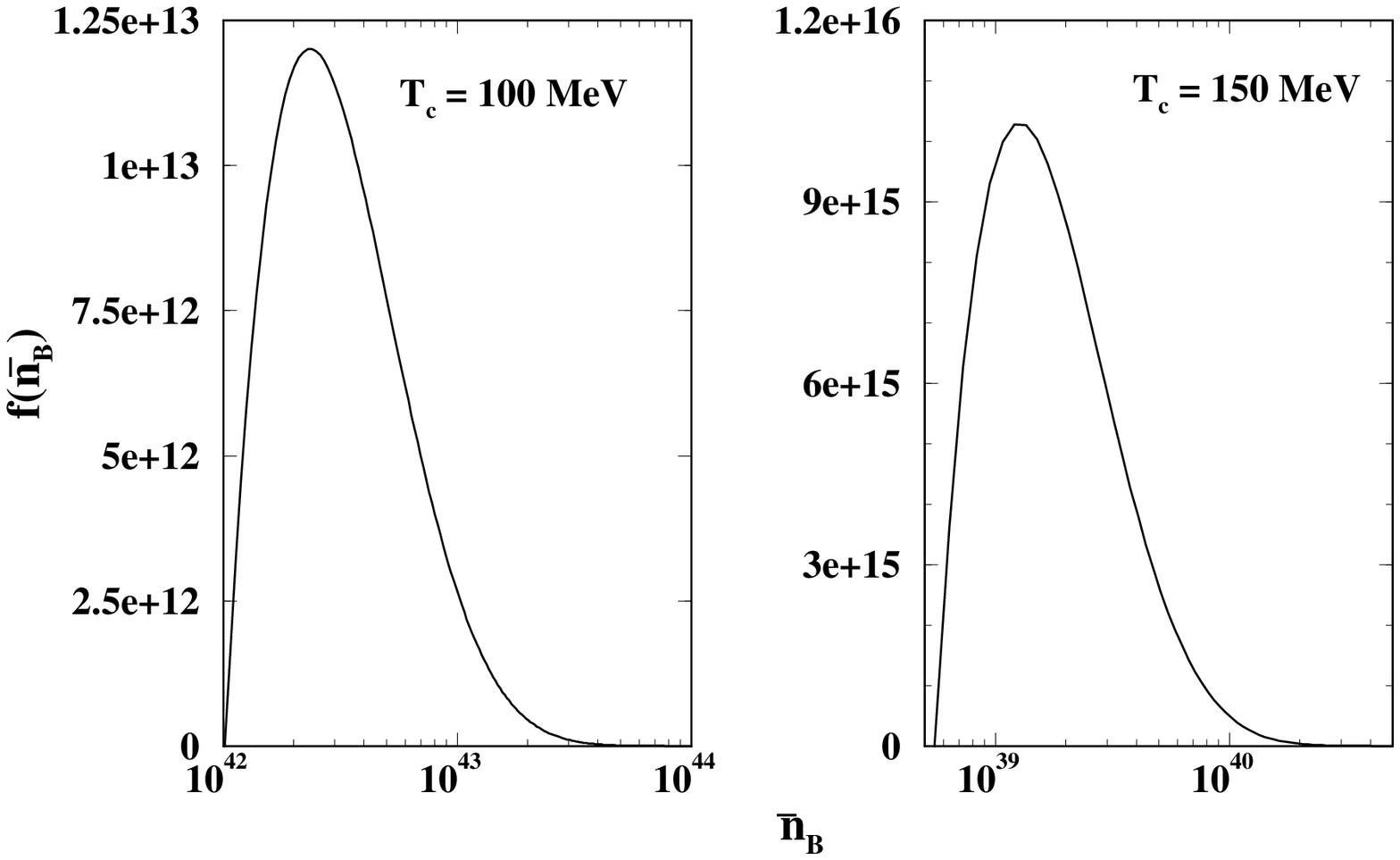,height=8cm,width=15cm}}
\caption{Same as fig.~\protect\ref{cot}, using nucleation rate proposed
by Csernai and Kapusta. The value of $\sigma$ is 
$10 MeV fm^{-2}$.}
\label{10kap}
\eef
\newpage
\bef[ht]
\centerline{\psfig{figure=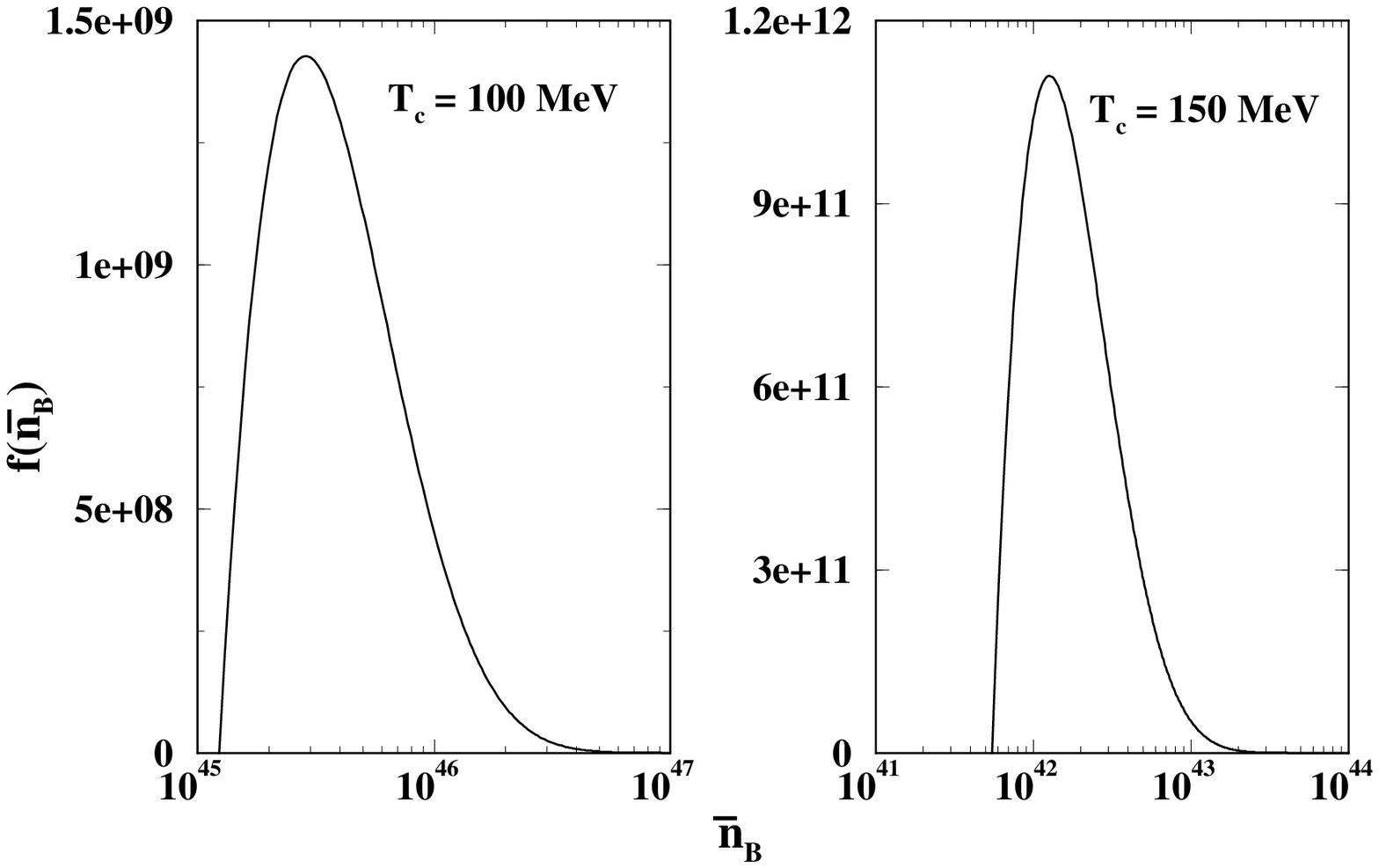,height=8cm,width=15cm}}
\caption{Same as fig.~\protect\ref{10kap} with $\sigma=50$ MeV fm$^{-2}$.}
\label{50kap}
\eef
\newpage
\bef[hb]
\centerline{\psfig{figure=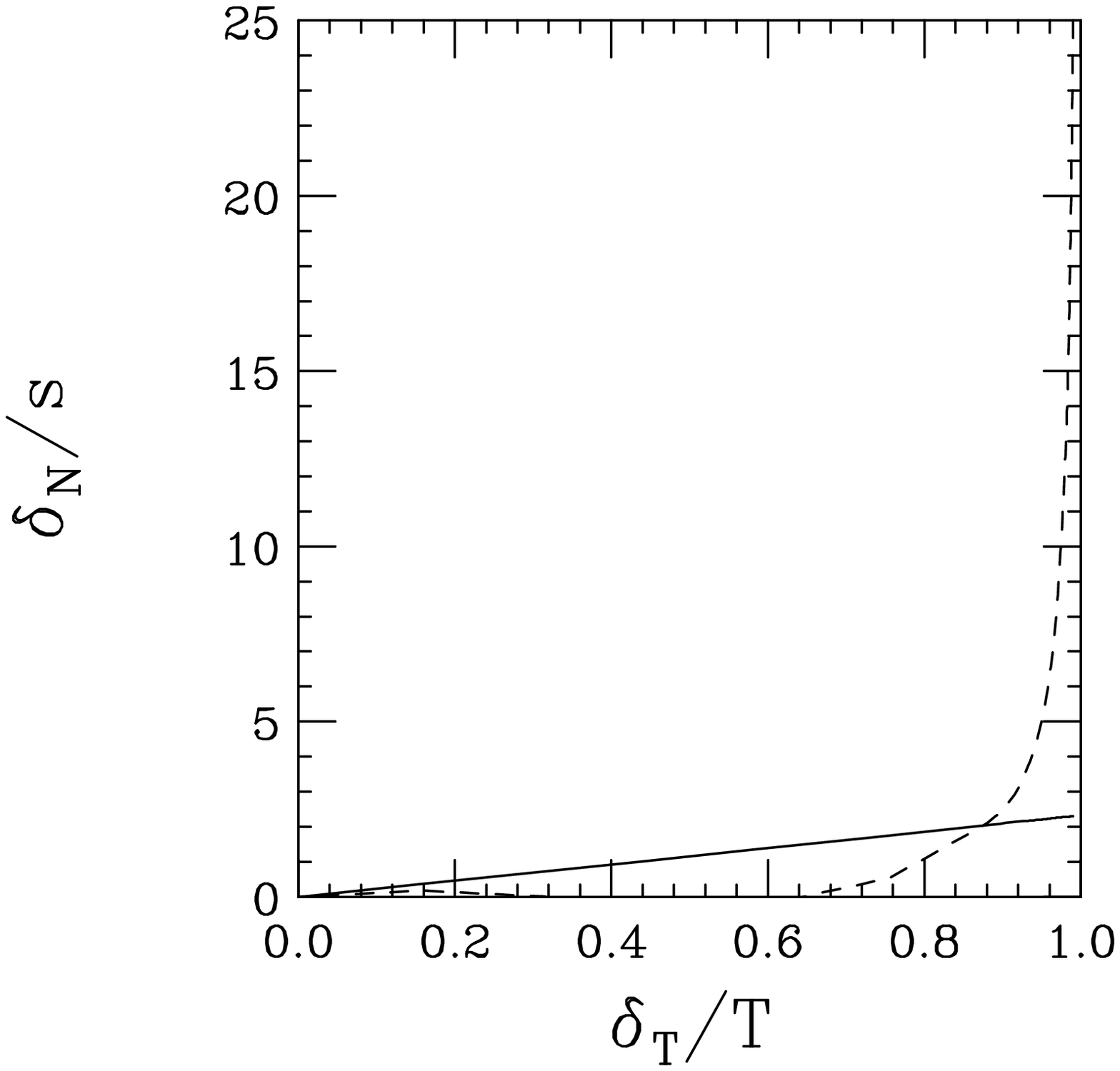,height=10cm,width=11cm}}
\caption{The relation between baryon inhomogeneity and temperature
difference. The dotted line corresponds to the solution of 
eq.~(\protect\ref{a1}) 
and the solid line corresponds to the linear approximation as discussed
in the text.}
\label{inhom}
\eef

\newpage
\bef[htb]
\centerline{\psfig{figure=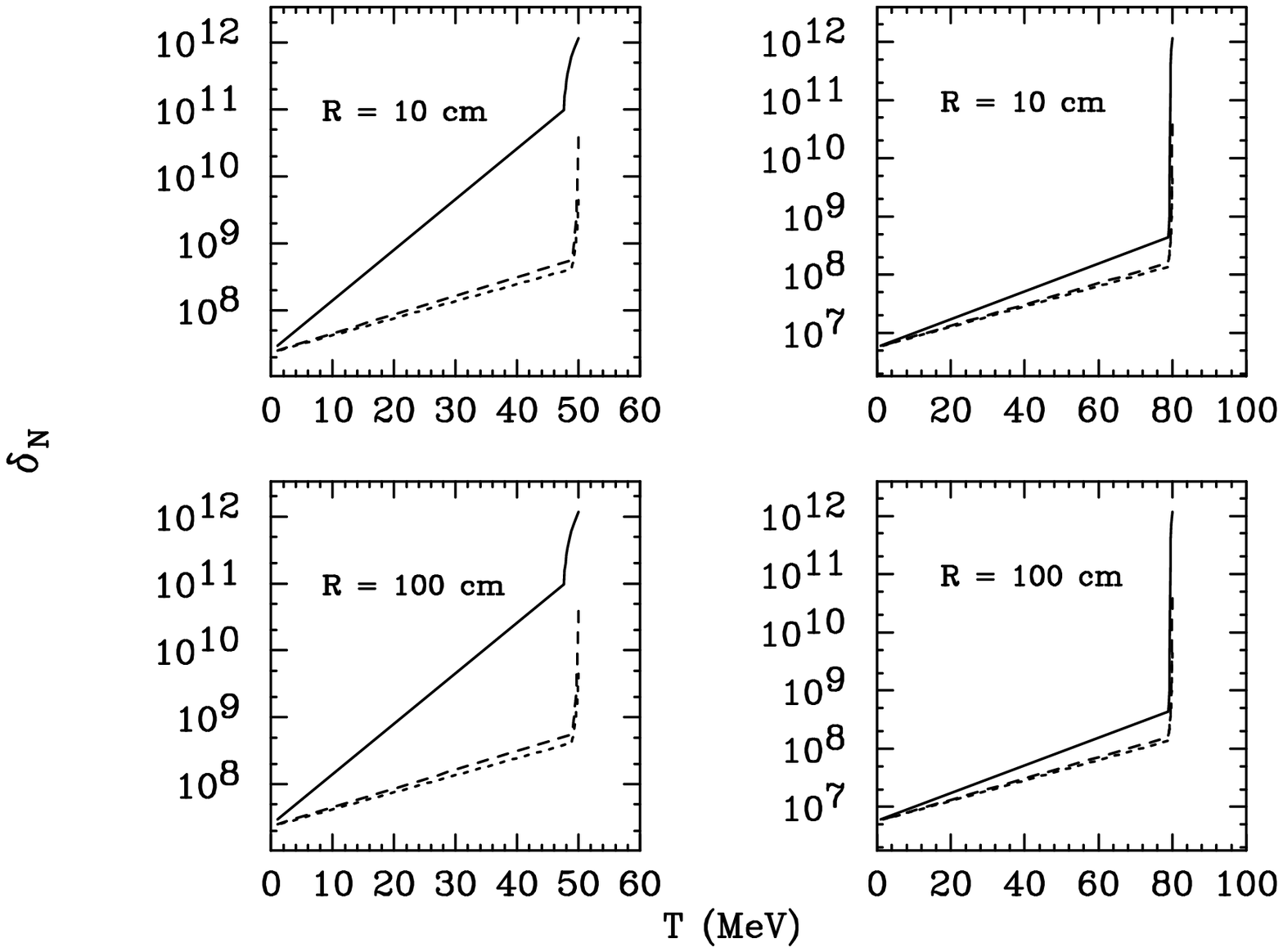,height=10cm,width=11cm}}
\caption{The evolution of baryon inhomogeneity with temperature, 
for different sizes of the clump and different initial temperature, due
to neutrino inflation.}
\label{infl}
\eef
\end{document}